\documentclass[trackchanges]{aastex701}
\usepackage{subcaption}

\usepackage{xcolor}
\usepackage{soul}

\begin{document}

\title{No Strong Evidence for Plasma Lensing in FRB 20240114A}

\author[orcid=0000-0001-8065-4191,gname=Jiarui,sname=Niu]{Jiarui Niu}
\affiliation{The Hong Kong Institute for Astronomy and Astrophysics, The University of Hong Kong, Pokfulam Road, Hong Kong, China}
\affiliation{Department of Physics, The University of Hong Kong, Pokfulam Road, Hong Kong, China}
\affiliation{State Key Laboratory of Radio Astronomy and Technology, National Astronomical Observatories, CAS, A20 Datun Road, Chaoyang District, Beijing, 100101, P. R. China}
\email[show]{niujr@hku.hk}

\author[orcid=0009-0004-7249-8060, gname=Nan,sname=Xu]{Nan Xu}
\affiliation{The Hong Kong Institute for Astronomy and Astrophysics, The University of Hong Kong, Pokfulam Road, Hong Kong, China}
\affiliation{Department of Physics, The University of Hong Kong, Pokfulam Road, Hong Kong, China}
\email{nan.xu.astro@connect.hku.hk}

\author[orcid=0000-0002-2552-7277,gname=Xiaohui,sname=Liu]{Xiaohui Liu}
\affiliation{State Key Laboratory of Radio Astronomy and Technology, National Astronomical Observatories, CAS, A20 Datun Road, Chaoyang District, Beijing, 100101, P. R. China}
\affiliation{School of Astronomy and Space Science, University of Chinese Academy of Sciences, Beijing 100049, People's Republic of China}
\email{liuxh@bao.ac.cn}

\author[orcid=0009-0000-6275-3452, gname=Songyu,sname=Shen]{Songyu Shen}
\affiliation{School of Physics and Astronomy, Beijing Normal University, Beijing 100875, China}
\affiliation{Institute for Frontiers in Astronomy and Astrophysics, Beijing Normal University, Beijing 102206, China}
\email{syshen@mail.bnu.edu.cn}

\author[orcid=0009-0005-8586-3001,gname=Junshuo,sname=Zhang]{Junshuo Zhang}
\affiliation{State Key Laboratory of Radio Astronomy and Technology, National Astronomical Observatories, CAS, A20 Datun Road, Chaoyang District, Beijing, 100101, P. R. China}
\affiliation{School of Astronomy and Space Science, University of Chinese Academy of Sciences, Beijing 100049, People's Republic of China}
\email{zhangjs@bao.ac.cn}

\author[orcid=0000-0002-9332-5562,gname=Tiancong,sname=Wang]{Tiancong Wang}
\affiliation{School of Physics and Astronomy, Beijing Normal University, Beijing 100875, China}
\affiliation{Institute for Frontiers in Astronomy and Astrophysics, Beijing Normal University, Beijing 102206, China}
\email{wangtc@mail.bnu.edu.cn}

\author[gname=Pawan,sname=Kumar]{Pawan Kumar}
\affiliation{Department of Astronomy, University of Texas at Austin, Austin, TX 78712, USA}
\email{pk@astro.as.utexas.edu}

\author[gname=Yuanhong,sname=Qu]{Yuanhong Qu}
\affiliation{Department of Physics, University of Helsinki, P.O. Box 64, FI-00014 University of Helsinki, Finland}
\email{yuanhong.y.qu@helsinki.fi}

\author[gname=Dejiang,sname=Zhou]{Dejiang Zhou}
\affiliation{State Key Laboratory of Radio Astronomy and Technology, National Astronomical Observatories, CAS, A20 Datun Road, Chaoyang District, Beijing, 100101, P. R. China}
\email{djzhou@nao.cas.cn}

\author[gname=Weiwei,sname=Zhu]{Weiwei Zhu}
\affiliation{State Key Laboratory of Radio Astronomy and Technology, National Astronomical Observatories, CAS, A20 Datun Road, Chaoyang District, Beijing, 100101, P. R. China}
\affiliation{Institute for Frontiers in Astronomy and Astrophysics, Beijing Normal University, Beijing 102206, China}
\email[show]{zhuww@nao.cas.cn}

\author[gname=Bing,sname=Zhang]{Bing Zhang}
\affiliation{The Hong Kong Institute for Astronomy and Astrophysics, The University of Hong Kong, Pokfulam Road, Hong Kong, China}
\affiliation{Department of Physics, The University of Hong Kong, Pokfulam Road, Hong Kong, China}
\email[show]{bzhang1@hku.hk}

\author[orcid=0000-0003-2516-6288, gname=He,sname=Gao]{He Gao}
\affiliation{School of Physics and Astronomy, Beijing Normal University, Beijing 100875, China}
\affiliation{Institute for Frontiers in Astronomy and Astrophysics, Beijing Normal University, Beijing 102206, China}
\email{gaohe@bnu.edu.cn}

\author[gname=Dongzi,sname=Li]{Dongzi Li}
\affiliation{Department of Astronomy, Tsinghua University, Beijing 100084, China.}
\email{dzli@mail.tsinghua.edu.cn}

\author[gname=Jinlin,sname=Han]{Jinlin Han}
\affiliation{State Key Laboratory of Radio Astronomy and Technology, National Astronomical Observatories, CAS, A20 Datun Road, Chaoyang District, Beijing, 100101, P. R. China}
\email{hjl@bao.ac.cn}

\author[gname=Di,sname=Li]{Di Li}
\affiliation{Department of Astronomy, Tsinghua University, Beijing 100084, China.}
\email{dili@mail.tsinghua.edu.cn}

\author[0000-0001-6475-8863]{Xuelei Chen}
\affiliation{State Key Laboratory of Radio Astronomy and Technology, National Astronomical Observatories, CAS, A20 Datun Road, Chaoyang District, Beijing, 100101, P. R. China}
\affiliation{School of Astronomy and Space Science, University of Chinese Academy of Sciences, Beijing 100049, People's Republic of China}
\email{xuelei@cosmology.bao.ac.cn}

\author[gname=Ke-Jia,sname=Lee]{Kejia Lee}
\affiliation{Department of Astronomy, School of Physics, Peking University, Beijing 100871, China}
\affiliation{National Astronomical Observatories, Chinese Academy of Sciences, Beijing 100101, China}
\affiliation{Yunnan Astronomical Observatories, Chinese Academy of Sciences, Kunming 650216, China}
\affiliation{Beijing Laser Acceleration Innovation Center, Huairou, Beijing 101400, China}
\email{kjlee@pku.edu.cn}

\author[orcid=0000-0001-5931-2381,gname=Ye,sname=Li]{Ye Li}
\affiliation{Purple Mountain Observatory, Chinese Academy of Sciences, Nanjing 210023, China}
\email{yeli@pmo.ac.cn}

\author[gname=Wei-Yang,sname=Wang]{Weiyang Wang}
\affiliation{School of Astronomy and Space Science, University of Chinese Academy of Sciences, Beijing 100049, People's Republic of China}
\email{wywang@ucas.ac.cn}

\author[gname=Qiuyang,sname=Fu]{Qiuyang Fu}
\affiliation{National Astronomical Observatories, Chinese Academy of Sciences, Beijing 100101, China}
\affiliation{School of Astronomy and Space Science, University of Chinese Academy of Sciences, Beijing 100049, People's Republic of China}
\email{fuqy@bao.ac.cn}

\author[gname=Jiawei,sname=Jin]{Jiawei Jin}
\affiliation{National Astronomical Observatories, Chinese Academy of Sciences, Beijing 100101, China}
\affiliation{School of Astronomy and Space Science, University of Chinese Academy of Sciences, Beijing 100049, People's Republic of China}
\email{jinjw@bao.ac.cn}

\author[gname=Yanqing,sname=Cai]{Yanqing Cai}
\affiliation{National Astronomical Observatories, Chinese Academy of Sciences, Beijing 100101, China}
\affiliation{School of Astronomy and Space Science, University of Chinese Academy of Sciences, Beijing 100049, People's Republic of China}
\email{caiyq@bao.ac.cn}

\author[gname=Caisong,sname=Liu]{Caisong Liu}
\affiliation{National Astronomical Observatories, Chinese Academy of Sciences, Beijing 100101, China}
\affiliation{School of Astronomy and Space Science, University of Chinese Academy of Sciences, Beijing 100049, People's Republic of China}
\email{liucs@bao.ac.cn}

\author[gname=Shuo,sname=Cao]{Shuo Cao}
\affiliation{National Astronomical Observatories, Chinese Academy of Sciences, Beijing 100101, China}
\email{caoshuo@bao.ac.cn}

\author[gname=Zi-Wei,sname=Wu]{Ziwei Wu}
\affiliation{National Astronomical Observatories, Chinese Academy of Sciences, Beijing 100101, China}
\email{wuzw@bao.ac.cn}

\author[gname=Dengke,sname=Zhou]{Dengke Zhou}
\affiliation{Research Center for Astronomical Computing, Zhejiang Laboratory, Hangzhou 311121, China}
\email{zdk@zhejianglab.com}

\author[orcid=0009-0002-3020-9123,gname=Long-Xuan,sname=Zhang]{Longxuan Zhang}
\affiliation{School of Physics, Huazhong University of Science and Technology, Wuhan 430074, China}
\email{d202580060@hust.edu.cn}

\author[gname=Wan-Jin,sname=Lu]{Wanjin Lu}
\affiliation{National Astronomical Observatories, Chinese Academy of Sciences, Beijing 100101, China}
\affiliation{School of Astronomy and Space Science, University of Chinese Academy of Sciences, Beijing 100049, People's Republic of China}
\email{wjlu@bao.ac.cn}

\author[gname=Yi,sname=Feng]{Yi Feng}
\affiliation{Research Center for Astronomical Computing, Zhejiang Laboratory, Hangzhou 311121, China}
\affiliation{Institute for Astronomy, School of Physics, Zhejiang University, Hangzhou 310027, China}
\email{yifeng@zhejianglab.org}

\author[gname=Chen-Hui,sname=Niu]{Chenhui Niu}
\affiliation{Institute of Astrophysics, Central China Normal University, Wuhan 430079, China}
\email{niuchenhui@ccnu.edu.cn}

\author[gname=Jia-Wei,sname=Luo]{Jiawei Luo}
\affiliation{College of Physics and Hebei Key Laboratory of Photophysics Research and Application, Hebei Normal University, Shijiazhuang 050024, China}
\affiliation{Shijiazhuang Key Laboratory of Astronomy and Space Science, Hebei Normal University, Shijiazhuang 050024, China}
\email{ljw@hebtu.edu.cn}

\author[orcid=0000-0002-4300-121X,gname=Rui,sname=Luo]{Rui Luo}
\affiliation{Department of Astronomy, School of Physics and Materials Science, Guangzhou University, Guangzhou 510006, China}
\email{rui.luo@gzhu.edu.cn}

\author[gname=Chun-Feng,sname=Zhang]{Chunfeng Zhang}
\affiliation{National Time Service Center, Chinese Academy of Sciences, Xi’an 710600, China}
\email{zhangchunfeng@ntsc.ac.cn}

\author[gname=Shiqian,sname=Zhao]{Shiqian Zhao}
\affiliation{Department of Astronomy, School of Physics and Materials Science, Guangzhou University, Guangzhou 510006, China}
\email{sqzhao@e.gzhu.edu.cn}

\author[gname=Chengwei,sname=Liang]{Chengwei Liang}
\affiliation{Department of Astronomy, School of Physics and Materials Science, Guangzhou University, Guangzhou 510006, China}
\email{rui.luo@gzhu.edu.cn}


\begin{abstract}
FRB~20240114A is an extremely active repeating fast radio burst for which plasma lensing has been proposed to explain its burst-rate variations, spectral evolution, and apparently ``carbon-copy'' burst pairs. Using FAST data and publicly available Parkes observations, we test this interpretation with the predictions of a one-dimensional Gaussian plasma-lens model. Although the burst-rate enhancements may be fitted separately for Parkes and FAST, the corresponding magnification peaks and demagnification troughs between the two datasets are offset from each other, inconsistent with plasma lensing predictions. The enhancement peaks also do not show consistent periodicity, disfavoring a binary plasma lensing origin. The burst bandwidths are not systematically narrower during the proposed lensing interval than in other episodes. Most importantly, the burst energy distribution, which shows an intrinsic turnover peak at low energies, {shows no significant shift of the turnover during the magnification interval, providing a model-independent test to the plasma-lensing interpretation.} All these suggest that there is no strong evidence in favor of a plasma lensing interpretation to the source. Rather, all the burst property variations are consistent with being caused by the intrinsic variations of the FRB central engine.

\end{abstract}


\section{Introduction}\label{sec:introduction}

Fast radio bursts (FRBs) are millisecond-duration radio transients of extragalactic origin \citep{2007Sci...318..777L}. Repeating FRBs are increasingly found to show multi-component burst morphologies, complex spectral behavior, and diverse polarization properties. These observations suggest that the detected burst properties reflect not only the physics of the emission process, but also propagation effects in the immediate environment of the source\citep{2019ARA&A..57..417C,2018Natur.553..182M,2020Natur.587...45Z, 2023RvMP...95c5005Z}.

FRB~20240114A was first discovered by the CHIME/FRB Collaboration and reported in 2024 January as a highly active repeating source \citep{2024ATelFRB240114A}. 
Its low declination gives CHIME only limited daily exposure to the source, so even the initial detections already implied an unusually high activity level \citep{2026ApJ...997..334S}. 
Following CHIME's report, the source was rapidly followed up by several radio telescopes \citep{2024ATel16430....1U,2024ATel16432....1O,2024ATel16433....1Z,2024ATel16446....1T}. 
Early MeerKAT observations detected 62 bursts from FRB~20240114A using the UHF and L band receivers and localized the source to a star-forming dwarf galaxy \citep{2024MNRAS.533.3174T}. 
These bursts are typically band-limited and exhibit high linear polarization fractions, suggesting a highly coherent emission process in a magnetized environment. 
The structure-optimizing dispersion measure (DM) was measured to be $527.65\pm0.01~\mathrm{pc~cm^{-3}}$ \citep{2024MNRAS.533.3174T}. 
Optical spectroscopic observations show that the host galaxy is a dwarf system at $z \approx 0.13$, with a stellar mass of $\sim 4\times10^8~M_\odot$ and ongoing star formation \citep{2025ApJ...980L..24C,2025ApJ...992L..35B}. 
This combination of a low-mass dwarf host and ongoing star formation is broadly consistent with the host environments of other highly active repeating FRBs, such as FRB~20121102A \citep{2017ApJ...834L...7T} and FRB~20190520B \citep{2022Natur.606..873N}. 
Complementary monitoring with small radio telescopes was also carried out, detecting 178 high-energy bursts over $\sim4200$ hr across 806 days with isotropic-equivalent energies of $\sim10^{40}$--$10^{42}$ erg \citep{2026arXiv260518513O}. 
Subsequent very long baseline interferometry observations revealed a compact persistent radio source (PRS) spatially associated with the burst location \citep{2024arXiv241201478B}. 
This association suggests that FRB~20240114A may reside in a dense and magnetized local environment, potentially linked to a young magnetar embedded in a nebula or supernova remnant.

The extreme activity of FRB~20240114A provides a unique opportunity to study both the burst emission process and propagation effects. 
Observations with the Five-hundred-meter Aperture Spherical radio Telescope (FAST) have detected more than $10^4$ bursts within several months, {revealing substantial variations in the detected burst rate between observing epochs, with the peak rate reaching about 700 bursts per hour} \citep{2025fastzhangjs,2026fastzhoudk,2026fastzhanglx,2026fastwangtc}. 
Such an extraordinary burst rate challenges standard magnetar emission models unless enhanced radio efficiency, strong beaming, or propagation-induced amplification effects are involved \citep{2025fastzhangjs}. 
Polarimetric monitoring further reveals rotation measure (RM) variations of order $\sim 200~\mathrm{rad~m^{-2}}$ over $\sim 200$ days, while the DM remains nearly constant, indicating a dynamically evolving magneto-ionic environment \citep{2026fastwangtc}. 
In addition to its high burst rate, FRB~20240114A shows pronounced spectral variability. 
Broadband observations between $\sim 1$--$6$~GHz reveal diverse burst morphologies, including narrowband emission, complex multi-component structures, and frequency-drifting features \citep{2025arXiv251008367L}. 
Observations with the Murriyang/Parkes ultra-wide bandwidth low receiver(UWL) further show long-term variations in the central emission frequency over months, as well as correlations on millisecond to minute timescales \citep{2026arXiv260216409U}. 
These results show that the observed phenomenology of FRB~20240114A reflects a combination of intrinsic source activity and propagation through a complex local environment.

{The observed long-term variations in the burst rate of FRBs may broadly arise from three classes of effects: observational selection effects, propagation-induced amplification or deamplification, and intrinsic variability of the source \citep{2019ARA&A..57..417C,2023RvMP...95c5005Z}. These possibilities affect the observed burst rate in different ways and therefore need to be considered separately.}

{First, observational selection effects influence both the measured burst rate and its temporal evolution. The number of bursts detected during a given epoch depends jointly on the source activity and the observational response, including the telescope sensitivity, observing exposure and cadence, frequency coverage, radio-frequency-interference environment and excision, detection threshold, and completeness of the burst-search pipeline \citep{2015MNRAS.447.2852K,2021ApJS..257...59C}. These effects are particularly relevant for FRBs because many bursts occupy only a limited frequency range, and their central emission frequencies evolve with time. Changes in the overlap between the burst spectrum and the observing band can therefore affect the measured rate, while differences in sensitivity and frequency coverage can lead different telescopes to sample different portions of the burst population.}
{Second, propagation through intervening plasma can amplify or attenuate burst signals. Such modulation changes the probability that individual bursts exceed the detection threshold and can therefore modify the observed burst rate relative to the rate and energy distribution at the source \citep{2017ApJ...842...35C,2018Natur.557..522M}.}
Plasma lensing provides one possible mechanism for such modulation.
It has long been discussed as a mechanism capable of modulating the apparent properties of compact radio sources. 
Early observational evidence came from extreme scattering events (ESEs), in which compact extragalactic radio sources showed strong, frequency-dependent flux density variations that were difficult to explain as intrinsic variability \citep{1987Natur.326..675F}. 
These events were soon interpreted as refractive focusing by compact ionized structures in the interstellar medium, capable of producing radio caustics and multiple imaging \citep{1987Natur.328..324R}. 
A later real-time detection of an ESE further demonstrated that plasma lensing can imprint frequency-dependent spectral structure during the event, providing observational constraints on Galactic plasma lenses \citep{2016Sci...351..354B}.
Subsequent theoretical work developed simple plasma-lens models, such as Gaussian electron-column-density lenses, showing that plasma refraction can produce highly chromatic magnification, demagnification, caustics, apparent position wander, and time-dependent light curves as the source, lens, and observer move relative to one another \citep{1998ApJ...496..253C}. 
In FRBs, plasma structures in the host galaxy or in the immediate source environment can produce strong magnification or demagnification, narrowband spectral structures, apparent changes in burst rate, and, in some cases, multiple imaging with arrival-time and DM perturbations \citep{2017ApJ...842...35C}. 
Several observational studies have discussed plasma lensing as a possible explanation for complex FRB phenomenology \citep{2019ApJ...876L..23H,2021MNRAS.505.3041P,2024MNRAS.531.4155C}. 
More recent theoretical work has further shown that plasma lensing can affect not only burst fluence and rate, but also sub-burst frequency drift, polarization angle jumps, and population luminosity functions \citep{2026ApJ...998..127L,2026ApJ..1001...28E}. 

Third, intrinsic source activity or changes in emission geometry may drive the evolution of the burst rate. In magnetar models, episodic magnetic energy release, crustal failure, starquakes, magnetic reconnection, and changes in magnetospheric twist, currents, or plasma loading can alter burst activity \citep{2019ApJ...879....4W,2023RvMP...95c5005Z}. Stochastic or avalanche like triggering can produce clusters of active episodes interspersed with quieter intervals. Some repeating FRBs show non Poissonian burst arrival statistics that support this possibility \citep{2018MNRAS.475.5109O}. Neutron star precession or orbital motion in a binary system can also change our view of the emission region and modulate the observed burst rate \citep{2020ApJ...892L..15Z,2020ApJ...893L..26I}. These processes can change how often the source produces bursts or how many bursts reach the observer. They may also affect burst energy, spectrum, morphology, and polarization.

For FRB~20240114A, plasma lensing has been proposed as one possible explanation for its observed spectral and rate variability. 
The Parkes observations in particular interpreted the burst-rate modulation, long-term central-frequency evolution, and repeated narrow-band spectral structures as possible lensing signatures \citep{2026arXiv260216409U}. 
The large FAST burst sample provides an independent and highly sensitive dataset {with which to test this interpretation and investigate alternative explanations for the observed rate variability}. 

The high sensitivity of FAST provides a large sample of bursts, while the UWL receiver of Parkes enables detailed characterization of spectral properties. 
By examining the evolution of burst rate, energy distribution, and fractional bandwidth ($\Delta\nu/\nu$), we aim to {test whether the burst-rate variations are accompanied by the energy and spectral changes expected from plasma-lensing magnification or demagnification, and to assess the relative roles of observational selection effects, propagation-induced modulation, and intrinsic source variability}.

\section{Observation}\label{sec:obs}

\begin{figure*}
	\includegraphics[width=18cm]{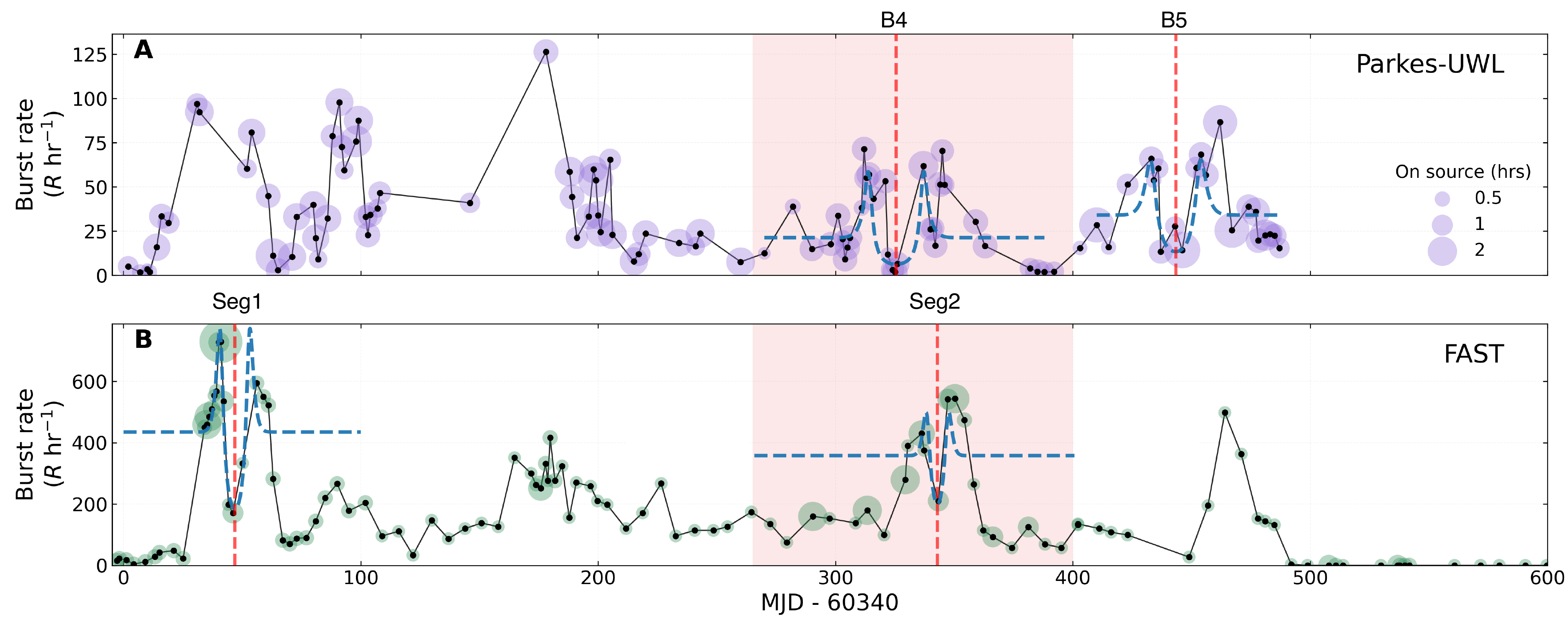}
    \caption{Burst-rate evolution of FRB~20240114A observed with Parkes and FAST. Panel A shows the burst rate measured with Parkes, where the marker size represents the on-source time (data taken from \citet{2026arXiv260216409U}), while Panel B shows the burst rate measured with FAST. The Parkes intervals B4 and B5 and the FAST intervals Seg1 and Seg2 indicate the time ranges used for the lensing analysis. B4 and Seg2, which cover approximately the same epoch, are highlighted in red for comparison. The blue dashed curves show the Gaussian-lens fits to the burst-rate evolution, and the red vertical lines mark the fitted centers of the corresponding attenuation troughs. Despite the temporal overlap between B4 and Seg2, the fitted trough centers are not aligned between the observations at different frequencies. No artificial bounds were imposed on the lens-model parameters during fitting, allowing both over-dense and under-dense lens solutions to be explored.
}
    \label{fig:rate}
\end{figure*}

FRB~20240114A has been intensively monitored with FAST since its discovery. The first FAST campaign covered 2024 January 28 to 2024 August 29, spanning 214 days with a total exposure time of 33.86 hr. Observations were conducted at L band (1.0--1.5 GHz), during which 11,553 bursts were detected above a fluence threshold of 0.026 Jy ms, with burst rates ranging from 4 to 729 hr$^{-1}$ and a mean rate of 249 hr$^{-1}$ \citep{2025fastzhangjs}. A subsequent campaign extended the monitoring baseline to 2025 May 30, covering 97 observing sessions with a total exposure time of 57.99 hr. The data were recorded in PSRFITS format \citep{Hotan_2004} with full-Stokes information, 4096 frequency channels, and a time resolution of 49.152~$\mu$s. Using a combination of \textsc{PRESTO} and machine-learning-assisted pipelines, a total of 17356 bursts with ${\rm S/N} \geq 12$ were identified, among which 6,131 bursts were selected for reliable polarimetric analysis. These observations reveal an extremely high burst rate and a dynamically evolving magneto-ionic environment \citep{2026fastwangtc}.

FRB~20240114A was also observed with Murriyang, the 64-m Parkes radio telescope, using the UWL receiver \citep{2026arXiv260216409U}. The Parkes campaign spanned approximately 16 months from MJD 60342 to 60827, with a total on-source time of 154 hr. The UWL receiver provides continuous frequency coverage from 704 to 4032 MHz, enabling broadband characterization of burst properties. The data were recorded in pulsar search mode, with a time resolution of 64~$\mu$s, and a frequency resolution of 0.5 MHz. To mitigate intra-channel dispersion smearing, coherent dedispersion was applied at a DM of 527.7~pc~cm$^{-3}$. A total of 5526 bursts were detected, corresponding to an average burst rate of $35.9 \pm 0.6$ hr$^{-1}$ \citep{2026arXiv260216409U}.

The results from FAST and Parkes provide complementary constraints on the burst rate, spectral properties, and temporal variability of FRB~20240114A, and offer an important observational basis for investigating the origin of its burst rate variability. {In this work, we compare the FAST and Parkes datasets to assess whether the observed burst rate evolution arises from selection effects associated with observations and data processing, propagation effects such as plasma lensing, intrinsic source variability, or a combination of these factors.} As shown in Figure~\ref{fig:rate}, the combined observational dataset is summarized in two panels. Panels A and B show the burst rate measured by Parkes and FAST, respectively, as a function of MJD, highlighting the strong temporal variability of the source.

\section{Result}\label{sec:res}
\subsection{Observation Selection Effects }
\begin{figure*}
	\includegraphics[width=18cm]{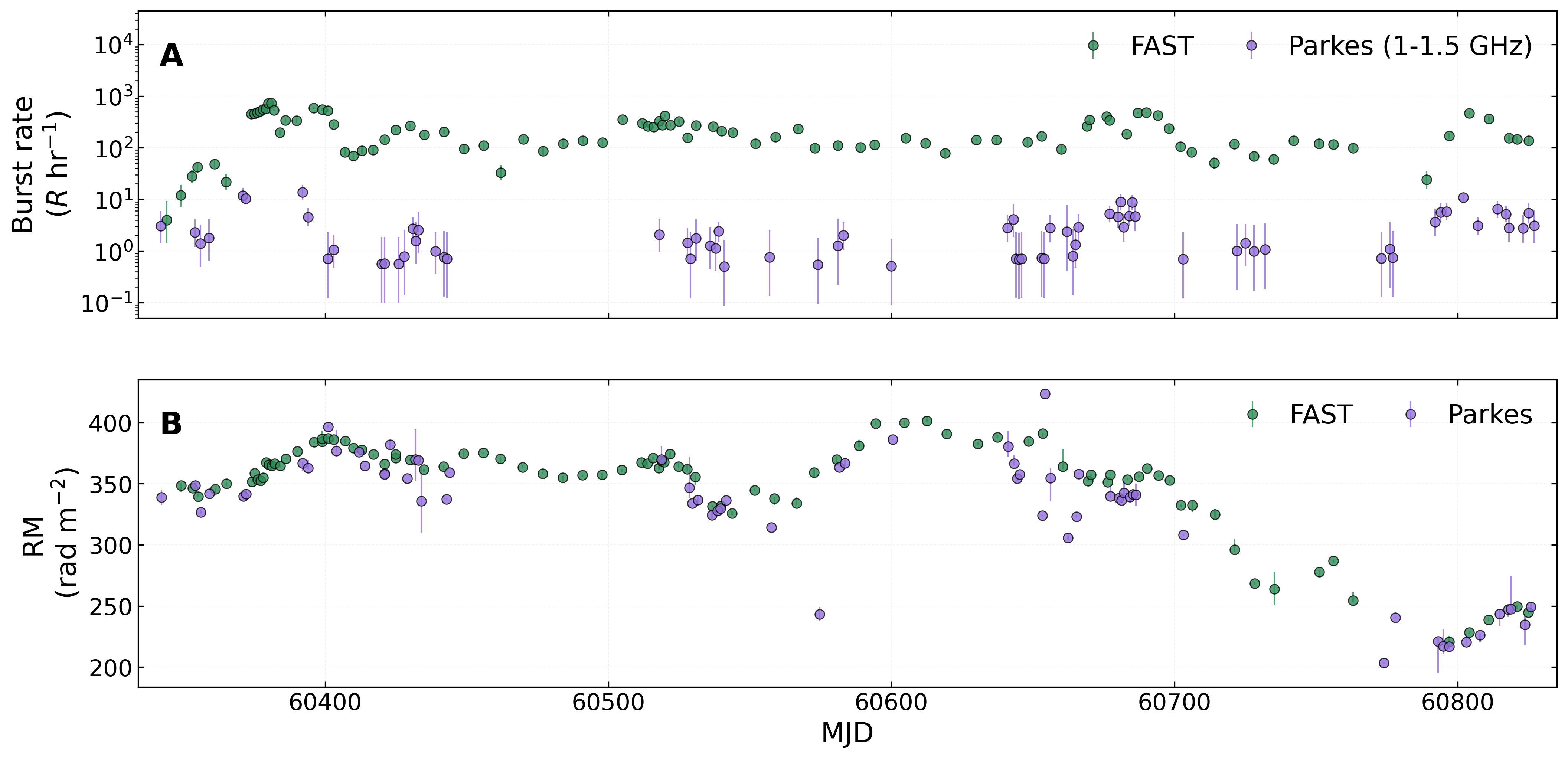}
    \caption{Burst-rate and RM evolution of FRB~20240114A observed with Parkes and FAST. Panel A presents the burst rates from Parkes and FAST, with the Parkes burst rate calculated using only bursts with central frequencies between 1.0 and 1.5 GHz (Parkes data taken from \citet{2026arXiv260216409U}). A logarithmic scale is used for the y-axis for comparison of the rate evolution. Panel B shows the RM measurements from FAST and Parkes.
}
    \label{fig:raterm}
\end{figure*}

{To examine whether observational selection effects can account for the burst rate evolution, we compare the FAST and Parkes measurements within their common frequency range from 1.0 to 1.5~GHz, as shown in Figure~\ref{fig:raterm}. Although the absolute rates differ because the telescopes have different sensitivities and detection thresholds, both datasets show broadly consistent temporal variations in burst activity. This agreement makes telescope response and data processing unlikely to be the primary causes of the observed rate evolution. The RM measurements from FAST and Parkes also follow a consistent evolution, showing that the magnetized plasma along the line of sight changed during the monitoring period. However, the burst rate shows no clear correlation with RM. This lack of correlation does not support a simple scenario in which the burst rate variations and RM evolution are produced by the same propagation structure, and is instead consistent with intrinsic variations in the source activity occurring alongside independent changes in the propagation environment.}

\subsection{One Dimensional Gaussian Plasma Lens}

\begin{figure*}
    \centering
    \includegraphics[width=18cm]{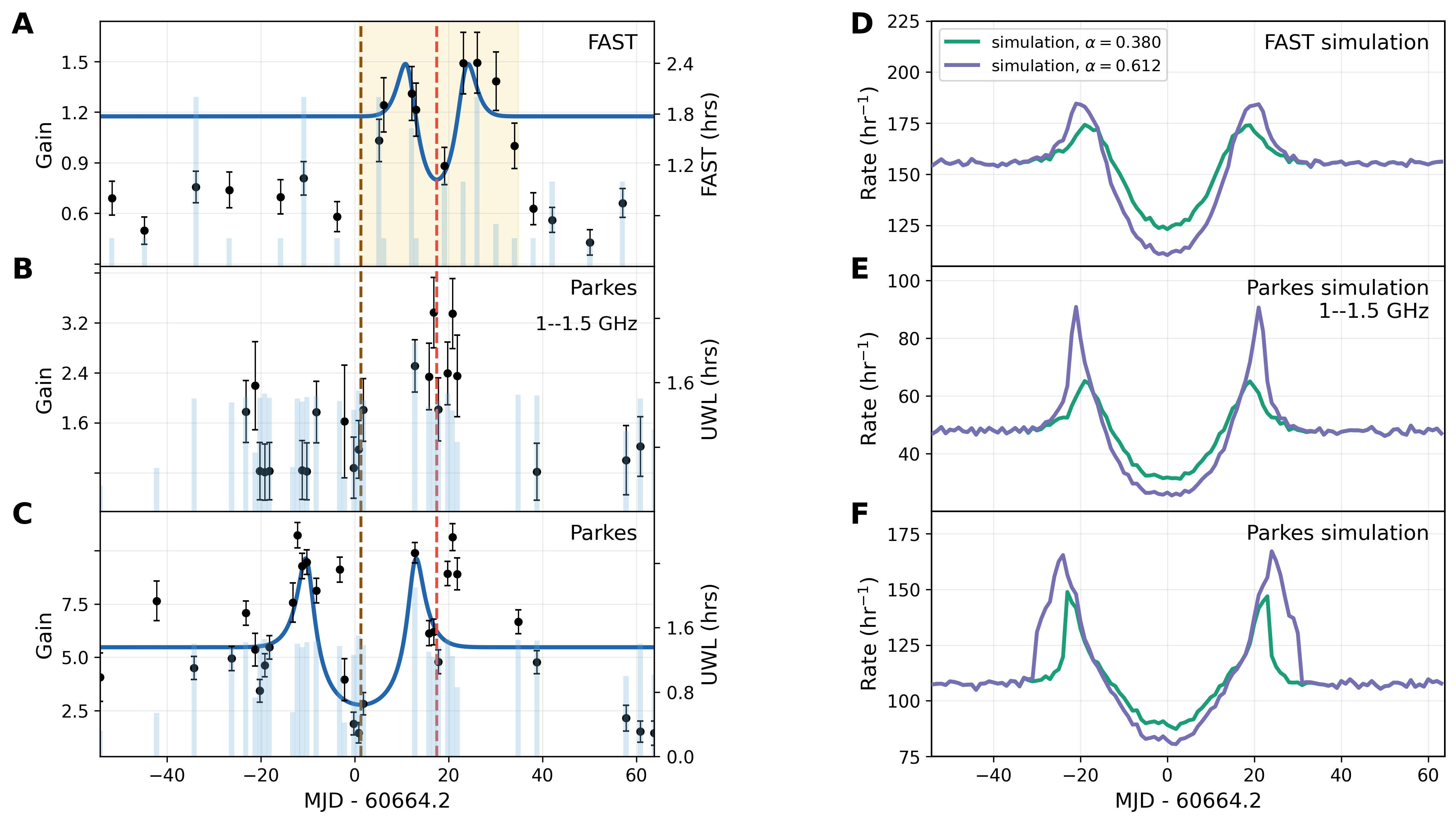}
    \caption{
Plasma lensing fits and simulations for the burst-rate modulation of FRB~20240114A. 
Panels A--C show the observational results over the same time interval. Panel A presents the FAST burst-rate variation together with the one-dimensional Gaussian plasma-lensing fit. Panel B shows the Parkes UWL observations using only bursts with central frequencies between 1.0 and 1.5 GHz, while Panel C shows the full-band Parkes UWL results together with the corresponding plasma-lensing fit. Panels D--F show the simulated burst-rate variations for FAST, Parkes 1.0--1.5 GHz, and the full Parkes UWL band, respectively, assuming a one-dimensional Gaussian plasma lens with lens strengths \(\alpha=0.380\) and \(0.612\). If the burst-rate variations observed by FAST and Parkes over the same time interval were produced by the same plasma lens, the expected behaviour would resemble the simulations: although the observing frequencies differ, the centers of the lensing-induced attenuation troughs remain aligned in time.}
    \label{fig:mcmc}
\end{figure*}

{Having examined observational selection effects, we next test the plasma lensing interpretation of the burst rate evolution.} Previous work \citep{2026arXiv260216409U} {modeled the Parkes burst rate variations using the magnification pattern of a one-dimensional Gaussian plasma lens. Here, we compare the FAST and Parkes observations to test whether this model can consistently account for the rate evolution measured by both telescopes. Figure~\ref{fig:rate} provides an overview of the burst-rate evolution observed with FAST and Parkes. Two prominent burst storms, B4 and B5, are identified in the Parkes observations, while two intervals, Seg1 and Seg2, are selected from the FAST data for comparison. In particular, Parkes B4 and FAST Seg2 cover approximately the same epoch, providing an opportunity to test whether the burst-rate variations observed by the two telescopes can be described by a common plasma lensing event.}

In this model, the observed burst-rate enhancement is converted into a proxy for the lensing gain by assuming that the intrinsic burst-fluence distribution follows a power law. For a differential fluence distribution
$dN/dF \propto F^{-\gamma}$, the cumulative burst rate above a fixed detection threshold scales with the lensing gain as
\[
R_{\rm obs} \propto G^{\gamma-1}.
\]
The corresponding observed gain proxy is therefore
\[
G_{\rm obs}(t)
=
\left[
\frac{R_{\rm obs}(t)}{R_0}
\right]^{1/(\gamma-1)},
\]
where $R_0$ denotes the unlensed baseline burst rate.

Following \citet{2026arXiv260216409U}, we adopt $\gamma=2.8$ and fit the resulting gain proxy using a one-dimensional Gaussian plasma-lens model. The lens equation is written as
\begin{equation}
    u' = u\left(1+\alpha e^{-u^2}\right),
\end{equation}
where $u$ and $u'$ are the dimensionless image-plane and source-plane coordinates, respectively. The dimensionless lens-strength parameter is defined as
\begin{equation}
    \alpha
    \equiv
    \frac{\lambda r_e R_F^2 N_0}
    {\pi a_{\rm lens}^2},
\end{equation}
where $\lambda$ is the observing wavelength, $r_e$ is the classical
electron radius, $R_F$ is the Fresnel radius, $N_0$ is the electron
column density at the center of the Gaussian lens, and $a_{\rm lens}$
is its characteristic transverse size
\citep{1998ApJ...496..253C,2017ApJ...842...35C}.
A larger value of $\alpha$ therefore corresponds to stronger plasma
refraction. For a given source-plane coordinate $u'$, the total
magnification is obtained by summing the magnifications of all image branches,
\begin{equation}
    G(u')
    =
    \sum_i
    \left|
    1+\alpha\left(1-2u_i^2\right)e^{-u_i^2}
    \right|^{-1},
\end{equation}
where each $u_i$ is a solution of the lens equation at the corresponding value of $u'$.

Assuming that the line of sight moves across the lens with an effective transverse velocity $v_{\rm trans}$, the source-plane coordinate evolves with time as
\begin{equation}
    u'(t)
    =
    \frac{
    v_{\rm trans}
    \left(
    t-t_{\rm ref}-t_{\rm shift}
    \right)
    }{
    a_{\rm lens}
    },
\end{equation}
where $a_{\rm lens}$ is the characteristic transverse size of the plasma lens. More precisely, it is the length scale over which the electron surface density in the lens plane decreases by a factor of $e$. The time-dependent gain is therefore modeled as
\begin{equation}
\label{eq:G_t}
    G_{\rm model}(t)
    =
    \mathcal{G}\,
    G\left[
    \frac{
    v_{\rm trans}
    \left(
    t-t_{\rm ref}-t_{\rm shift}
    \right)
    }{
    a_{\rm lens}
    }
    \right],
\end{equation}
where $\mathcal{G}$ is an overall gain normalization, $t_{\rm ref}$ is the reference epoch, and $t_{\rm shift}$ allows the center of the modeled lensing feature to shift relative to $t_{\rm ref}$. Because the temporal profile constrains only the ratio $v_{\rm trans}/a_{\rm lens}$, rather than $v_{\rm trans}$ and $a_{\rm lens}$ independently, the fitted parameters are $\alpha$, $v_{\rm trans}/a_{\rm lens}$, $t_{\rm shift}$, and $\mathcal{G}$. Equivalently, the ratio may be expressed in terms of an effective lens-crossing timescale,
\[
t_{\rm cross}
=
\frac{a_{\rm lens}}{v_{\rm trans}}.
\]
This approach assumes that the burst-rate variations within each selected storm are dominated by propagation-induced magnification rather than by intrinsic changes in the source activity.

{The overall burst-rate evolution and the corresponding plasma-lensing fits for the Parkes B4 and B5 intervals and the FAST Seg1 and Seg2 intervals are shown in Figure~\ref{fig:rate}. In particular, Parkes B4 and FAST Seg2 overlap in time, allowing the lensing interpretation to be tested directly using observations at different frequencies. We therefore focus on this common time interval in Figure~\ref{fig:mcmc}, where the observations are shown on an expanded timescale and compared with the expected frequency-dependent behaviour of the plasma-lensing model.  Panels A--C show the observations over the same time interval: Panel A presents the FAST burst-rate variation together with the plasma-lensing fit, Panel B shows the Parkes UWL observations restricted to bursts with central frequencies between 1.0 and 1.5~GHz, and Panel C shows the full-band Parkes UWL observations together with the corresponding lensing fit. We explored both over-dense and under-dense Gaussian plasma-lens solutions. The observed rise--dip--rise morphology is more naturally reproduced by the over-dense case, in which a central demagnification trough is bracketed by two magnification peaks.}

A more detailed examination reveals several inconsistencies that challenge a straightforward lensing interpretation.
First, we distinguish four quantities in the lensing interpretation. The \emph{fitted unlensed region} refers to the baseline burst rate inferred by the Gaussian lensing model in the absence of lensing. The \emph{observed unlensed region} refers to the burst rate actually measured before and after the candidate lensing interval. The \emph{fitted magnification region} is the part of the model where the plasma lens enhances the burst rate, while the \emph{fitted demagnification region}, or fitted trough, is the part where the lens suppresses the burst rate through defocusing. With these definitions, two inconsistencies become apparent.
First, the observed unlensed rate is systematically lower than the fitted unlensed rate in all candidate lensing intervals, particularly in the FAST observations. Therefore, reproducing the enhanced burst rate requires the model to assume an intrinsic baseline activity level that is higher than the rate actually observed outside the candidate lensing episode. Second, in several cases, the fitted demagnification trough remains above the observed unlensed rate. Because the trough represents a lens-suppressed state, it should lie below the true unlensed baseline. Instead, the burst rate observed before and after the storm is even lower than the fitted trough. The model therefore cannot simultaneously reproduce the magnified peak, the demagnified trough, and the observed off-storm activity level.

Second, the candidate lensing events in FAST Seg2 and Parkes B4 are not temporally aligned. {As shown in Panels A and C of Figure~\ref{fig:mcmc}, the fitted demagnification troughs occur at substantially different epochs despite the overlap in observing time. Panel B provides a frequency-matched comparison using only Parkes UWL bursts with central frequencies between 1.0 and 1.5~GHz, but these data do not show a corresponding trough at the epoch inferred from the FAST fit. To illustrate the behaviour expected if the FAST and Parkes burst-rate variations were produced by the same plasma lens, Panels D--F show simulated burst-rate curves for the FAST band, the Parkes 1.0--1.5~GHz range, and the full Parkes UWL band, respectively. We adopt the two lens strengths, $\alpha=0.380$ and $0.612$, inferred from the Parkes lensing fits; details of the simulations are given in the Appendix~\ref{app:forward_modeling}. Although the amplitudes and detailed shapes of the simulated magnification patterns vary with observing frequency, the centers of the demagnification troughs remain aligned in time. The temporal offset seen in the observations therefore cannot be reproduced simply by the chromatic response of the same one-dimensional Gaussian plasma lens, disfavouring a single coherent lens as a common explanation of the FAST and Parkes rate variations.}

Finally, {the specific lens fits assume that the} intrinsic burst rate of the source remains constant over the selected time intervals, and all observed variability is due to external magnification.
However, this assumption is not independently established by the observations and may not be appropriate for a highly active repeater such as FRB~20240114A, whose intrinsic activity can vary with time.
If the intrinsic burst rate evolves during the fitted interval, the observed rate modulation cannot be uniquely mapped to the lensing gain, and intrinsic variability may contribute to the apparent rise--dip--rise structure. The inferred lensing parameters and unlensed baseline should therefore be interpreted within this assumption.
The simulations in Panels D--F adopt the same constant intrinsic burst rate assumed in the lens fits, as their purpose is to illustrate the frequency-dependent behaviour expected from a single plasma lens under this assumption.

Overall, although individual burst storms can be phenomenologically fitted with a {one-dimensional Gaussian plasma-lens}
model, the inferred intrinsic baseline rate is higher than the burst rate observed outside the candidate lensing intervals. In several cases, the observed pre- and post-storm rates are even lower than the fitted demagnification trough.

Together with the different fitted epochs in the FAST and Parkes observations, and with the simulations showing that frequency-dependent lensing alone does not shift the demagnification trough to different epochs, these tensions reveal limitations of a one-dimensional Gaussian plasma lens with a constant intrinsic baseline as a common description of the two datasets. They do not rule out a contribution from plasma lensing. More complex lens geometries, multiple plasma structures, or models that allow simultaneous intrinsic variability could potentially alleviate these tensions. However, increasingly complex configurations may also weaken the specific observational predictions that can be tested against the data.

\subsection{Energy Distribution and Spectral Squeezing}
{Given the limitations of the one-dimensional Gaussian lens model discussed above, and because more complex plasma lens configurations remain possible, we next test for lensing signatures without adopting a specific lens model. If plasma lensing dominates the sharp increase in burst rate, its magnification should leave corresponding signatures in the burst energy and spectral properties. We therefore compare the energy distribution and fractional bandwidth ($\Delta\nu/\nu$) between the high and low activity intervals.}

Under the plasma lensing hypothesis, the observed burst energy is amplified by a frequency-dependent magnification factor $G$. For a burst population with an intrinsic power-law energy distribution, lensing can magnify intrinsically faint bursts that would otherwise fall below the detection threshold, bringing them into the observed sample \citep{2017ApJ...842...35C, 2026ApJ..1001...28E}. As a result, the cumulative energy distribution in a lensed interval is expected to be systematically elevated relative to an unlensed part. In addition, plasma lensing can introduce strongly frequency-dependent magnification. 
Near caustics, the lensing gain is not uniform across the observing band, but is instead enhanced only over a limited range of frequencies. As a result, even if the intrinsic burst emission is broader in frequency, only the strongly magnified part of the spectrum may rise above the detection threshold. This effect can make the observed emission appear more band-limited, effectively reducing the measured spectral occupancy and producing smaller values of $\Delta \nu / \nu$ \citep{2017ApJ...842...35C,2024ApJ...974..160K,2023ApJ...956...67Y}.

\begin{figure*}
	\includegraphics[width=18cm]{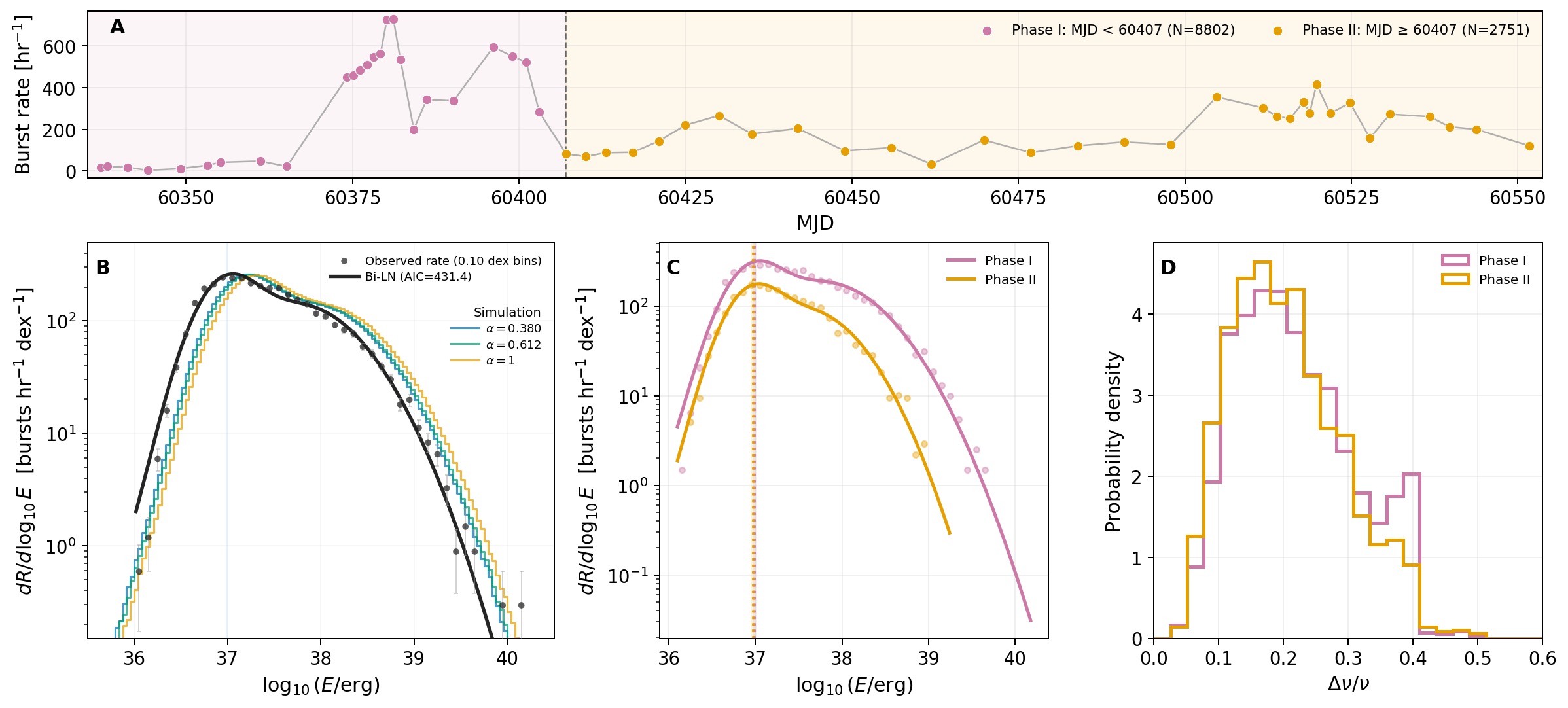}
    \caption{Energy function and fractional-bandwidth ($\Delta\nu/\nu$) statistics of FRB~20240114A based on the FAST observations. 
Panel A: Burst-rate evolution divided at MJD~60407 into Phase~I (pink) and Phase~II (orange). Phase~I contains the major high-rate burst episode, while Phase~II provides a lower-rate reference interval. 
Panel B: Energy function of all FAST bursts shown in Panel~A. The black points show the observed energy function, and the black solid curve shows the empirical bi-lognormal fit. The colored curves show simulated energy functions after plasma-lens magnification for $\alpha=0.380$, $0.612$, and $1$, starting from the intrinsic energy distribution. In these simulations, magnification shifts the turnover of the energy function toward higher energies. Simulation details are given in the Appendix~\ref{app:forward_modeling}. {Here, the one-dimensional Gaussian lens model is used only as an illustrative example. More generally, if plasma lensing magnifies the observed bursts, their apparent energies will be shifted accordingly, modifying the observed energy distribution. Therefore, a change in the characteristic energy scale is a generic consequence of lensing magnification and does not depend on the specific lens model adopted here.}
Panel C: Observed energy functions in Phase~I and Phase~II. Although Phase~I has a higher normalization because of its enhanced burst rate, no comparable shift of the turnover toward higher energies is apparent. 
Panel D: Fractional-bandwidth distributions for the same two phases. The distributions largely overlap, with no clear systematic shift or narrowing during the high-rate interval.
}
    \label{fig:energy_function}
\end{figure*}

{To further test whether the sharp increase in burst rate is dominated by plasma lensing, we examine the burst energy and bandwidth distributions of the FAST bursts. As shown in panel A of Figure~\ref{fig:energy_function}, we divide the FAST observations into two time intervals at MJD~60407. Phase~I corresponds to the interval with the highest burst activity, while Phase~II provides a lower-rate and comparatively more stable reference interval. This division is based only on the observed burst-rate behaviour and does not assume a specific lens model. A weak lensing contribution could in principle persist throughout the observations. But if the exceptionally high burst rate in Phase~I were mainly driven by strong lensing magnification, then the data should show corresponding observational signatures. 

The burst-energy distribution provides a direct test. The black solid line in Panel B shows the lognormal energy distribution of all FAST bursts included in Panel A, consistent with the treatment adopted by \citet{2025fastzhangjs} for FRB~20240114A. In our analysis, the black points show the observed differential energy distribution, and the black curve gives an empirical bi-lognormal fit that describes both the low-energy turnover and the high-energy tail. Using an intrinsic lognormal-like burst-energy distribution, we then simulate the effect of plasma-lens magnification for representative lens strengths $\alpha=0.380$, $0.612$, and $1$, with the simulation details given in the Appendix~\ref{app:forward_modeling}. In these simulations, lensing magnification shifts the burst energies to higher values and correspondingly moves the turnover of the energy distribution toward the high-energy side. {More generally, such a shift is expected when the observed rate enhancement is dominated by lensing magnification, rather than being specific to the Gaussian lens model.}  Therefore, if the enhanced burst rate during Phase~I were primarily caused by strong lensing, a similar shift of the turnover energy would be expected in the observed Phase~I distribution relative to Phase~II. Panel C compares the observed energy distributions of the two phases. The Phase~I distribution is higher in normalization, reflecting its larger burst rate, but it does not show the clear shift of the turnover toward higher energies predicted by the lensing simulations. The main difference between the two phases is therefore a change in burst occurrence rate, rather than the horizontal displacement of the characteristic energy scale expected if strong magnification dominated the burst rate enhancement. 

We also compare the fractional bandwidth distributions of the two phases in panel D. This provides a complementary test for chromatic propagation effects, because strong plasma lensing may also modify the observed spectral width and produce a systematic change in $\Delta\nu/\nu$. The two distributions largely overlap, with no obvious narrowing, broadening, or systematic displacement during the high-rate interval. We therefore do not find a distinct narrow-band signature associated with the burst storm. 

These tests do not rely on adopting a particular plasma-density profile for the lens. Although weak lensing effects cannot be excluded, the FAST data show no clear signatures expected if the high burst rate interval were dominated by strong lensing magnification. In particular, we do not observe the expected shift in the energy function turnover, providing a more general challenge to the lensing interpretation, while the fractional bandwidth distribution also shows no distinct change. The observations are instead more naturally explained by a substantial increase in the burst occurrence rate while the characteristic burst-energy scale remains approximately stable.}

\section{Conclusion}\label{sec:con}
The results presented above allow us to {assess the origin of the burst rate evolution of} FRB~20240114A. {We examine whether the observed changes can be attributed to telescope and data processing effects, propagation through plasma structures, or intrinsic source variability.}

First, the FAST and Parkes measurements within their common frequency range from 1.0 to 1.5~GHz show broadly consistent changes in burst activity despite the different telescope sensitivities and data processing procedures. This agreement makes instrumental and processing effects unlikely to be the primary cause of the observed rate evolution. The RM measurements from the two telescopes also follow a consistent evolution, confirming that the magnetized plasma along the line of sight changed during the monitoring period. However, the burst rate shows no clear correspondence with RM, providing no evidence that the same plasma structure controlled both the RM evolution and the burst storm.

Second, the burst-rate modelling reveals that individual burst storms (B4, B5, Seg1, and Seg2) can be fitted by a one-dimensional Gaussian plasma lens model. The fitted lensing parameters vary among different observing campaigns, which may be expected if distinct small plasma lenses are responsible for different episodes. We therefore do not regard the lack of parameter coherence across widely separated epochs as strong evidence against plasma lensing. However, even during overlapping epochs (e.g., B4 and Seg2), the candidate lensing events identified in the Parkes and FAST datasets are not temporally aligned, and the inferred lensing centers and trough positions are significantly offset. In addition, in several candidate intervals, particularly in the more sensitive FAST observations, the {observed unlensed rate} is lower than the fitted demagnification trough. Under the constant baseline assumption of the model, the trough should lie below the fitted unlensed baseline. Its position above the observed off storm rate shows that the inferred baseline does not consistently represent the surrounding activity. Together with the temporal offset between the two telescopes, these results show that a one-dimensional Gaussian plasma lens does not provide a common description of the FAST and Parkes datasets. More complex lens geometries, multiple plasma structures, or stronger chromatic effects could potentially alleviate these tensions. However, sufficiently complex configurations may be difficult to model and may not yield specific predictions that can be tested observationally.

Third, population-level tests based on energy and spectral properties do not reveal clear signatures of lensing. We compare the energy functions of the high activity Phase~I and the lower and more stable Phase~II. If magnification dominates the increase in burst rate, the turnover energy in Phase~I should shift toward higher energies. Instead, the Phase~I energy function increases mainly in normalization, while its turnover remains nearly aligned with that of Phase~II. This behavior is consistent with an increase in the number of bursts while the characteristic energy scale remains approximately stable. Similarly, {the fractional bandwidth distributions} ($\Delta \nu / \nu$) remain similar between the two phases, with no clear systematic narrowing, broadening, or displacement during the burst storm. The absence of a turnover shift and the similar fractional bandwidth distributions provide no independent evidence that plasma lensing dominates the large increase in burst rate.

These findings suggest that plasma lensing is not uniquely required to explain the observed phenomenology of FRB~20240114A. Although a one-dimensional Gaussian lens can provide a phenomenological description of individual burst storms, it does not consistently reproduce the FAST and Parkes rate evolution, and the energy and spectral comparisons do not reveal the accompanying signatures expected if magnification were the dominant cause. These results do not exclude more complex plasma lensing effects, but such effects are not required by the current observations.

The more natural interpretation is that the temporal evolution of the burst rate primarily reflects intrinsic changes in the source, such as variations in the magnetospheric state, emission geometry, or burst triggering process. The evolving propagation environment may independently affect the RM, polarization, or properties of individual bursts without controlling the overall burst rate. Plasma lensing may therefore contribute to some features, but intrinsic source activity provides the most consistent explanation for the observed rate evolution.

\begin{acknowledgments}
We thank the staff of the FAST telescope for their support during the observations. FAST is a Chinese national
mega-science facility, built and operated by the National Astronomical Observatories, Chinese Academy of Sciences.
This work made use of the data from FAST FRB Key Science Project.
This work is supported by the National Natural Science Foundation of China (NSFC, No. 12503055) and the Postdoctoral Fellowship Program of CPSF under Grant Number GZB20250737. YF is supported by National Natural Science Foundation of China grant No. 12522305. Y.Q. is supported by the Research Council of Finland Centre of Excellence in Neutron-Star Physics (project 374063).
\end{acknowledgments}

\appendix

\section{Forward Modeling of Plasma-Lensing Signatures}
\label{app:forward_modeling}

We use Monte Carlo simulations to construct the plasma-lensing expectations shown together with the observations in the main text. 
The same basic framework is used for both the burst-rate and energy-distribution analyses.
Synthetic bursts are first generated with prescribed energies and time--frequency morphologies and are then propagated through the frequency-dependent magnification pattern of the one-dimensional Gaussian plasma lens. 
Instrumental selection is included only when modeling the
detected burst rate.

\subsection{Burst-Rate Modeling}
\label{app:rate_modeling}

We generate a synthetic burst population with a constant mean intrinsic occurrence rate, with individual burst arrival times following a Poisson process. 
Each burst is assigned an equivalent isotropic input energy $E_{{\rm in},i}$ and an intrinsic time--frequency morphology before any lensing or instrumental effects are applied.

For the $i$th burst, we write the intrinsic time--frequency energy distribution as
\begin{equation}
    \mathcal{E}_{{\rm in},i}(\nu,\tau)
    =
    E_{{\rm in},i} f_i(\nu,\tau),
    \label{eq:input_waterfall}
\end{equation}
where $\nu$ is frequency, $\tau$ is time within an individual burst, and $\mathcal{E}_{{\rm in},i}$ has units of $\rm erg\,Hz^{-1}\,s^{-1}$.
The burst profile is normalized in the full time--frequency plane,
\begin{equation}
    \int\!\!\int
    f_i(\nu,\tau)\,d\nu\,d\tau
    =
    1,
    \label{eq:waterfall_normalization}
\end{equation}
so that integration over the intrinsic waterfall recovers $E_{{\rm in},i}$.

The formalism does not require a particular functional form for $f_i(\nu,\tau)$. 
In the present simulations, we adopt the simplest separable two-dimensional Gaussian profile,
\begin{equation}
    f_i(\nu,\tau)
    \propto
    \exp\left[
        -\frac{(\nu-\nu_{c,i})^2}{2\sigma_{\nu,i}^2}
        -\frac{(\tau-\tau_{c,i})^2}{2\sigma_{\tau,i}^2}
    \right],
    \label{eq:2d_gaussian}
\end{equation}
with the central frequency and spectral and temporal widths sampled independently from the adopted distributions.

For a burst arriving at observing epoch $t_i$, the lens modifies its time--frequency distribution according to
\begin{equation}
    \mathcal{E}_{{\rm lens},i}(\nu,\tau;t_i)
    =
    \mathcal{E}_{{\rm in},i}(\nu,\tau)
    G[\nu,u(t_i)],
    \label{eq:lensed_waterfall}
\end{equation}
where $G(\nu,u)$ is the frequency-dependent plasma-lensing gain and $u(t)$ describes the motion of the line of sight across the lens. 
Since the lens position changes negligibly during the duration of an individual burst, the gain depends on observing epoch and frequency but not on the intra-burst time $\tau$.

The total isotropic-equivalent energy after lensing is obtained by integrating the lensed waterfall,
\begin{equation}
    E_{{\rm lens},i}
    =
    \int\!\!\int
    \mathcal{E}_{{\rm lens},i}(\nu,\tau;t_i)
    \,d\nu\,d\tau .
    \label{eq:lensed_energy}
\end{equation}
Following the treatment used in the original simulation framework, this energy is converted to the observer-frame energy fluence as
\begin{equation}
    \mathcal{F}_{{\rm obs},i}
    =
    E_{{\rm lens},i}
    \frac{1+z}{4\pi D_L^2},
    \label{eq:observed_fluence}
\end{equation}
where $\mathcal{F}_{\rm obs}$ is in units of $\rm erg\,cm^{-2}$, $z$ is the source redshift, and $D_L$ is the luminosity distance.

The corresponding lensed time--frequency signal is then restricted to the observing band of each telescope and converted to the fluence convention used for the observations. 
Bursts below the corresponding detection threshold are removed, and the surviving events are used to construct the simulated FAST, frequency-matched Parkes, and full-band Parkes burst-rate curves shown in Figure~\ref{fig:mcmc}.

The intrinsic burst rate is held constant in these simulations so that the resulting temporal variation isolates the modulation produced by the plasma
lens.

\subsection{Energy-Distribution Modeling}
\label{app:energy_modeling}

The energy-distribution calculation uses the same burst-generation and plasma-lensing procedure described above. 
The input burst energies follow the empirical bi-lognormal energy distribution adopted in Figure~\ref{fig:energy_function}, and each burst is independently assigned a time--frequency morphology through $f_i(\nu,\tau)$ before propagation
through the lens.

To examine the energy redistribution produced by plasma-lens magnification, we select the simulated bursts within the lensing-induced rate-enhancement interval, where the lens is expected to have the strongest effect on the burst energies. 
The post-lensing energy of each selected burst is calculated directly from its lensed waterfall using Equation~(\ref{eq:lensed_energy}). Equivalently,
\begin{equation}
    E_{{\rm lens},i}
    =
    E_{{\rm in},i}
    \int\!\!\int
    f_i(\nu,\tau)
    G[\nu,u(t_i)]
    \,d\nu\,d\tau .
    \label{eq:lensed_energy_distribution}
\end{equation}

Because the plasma-lensing gain is chromatic and individual bursts have different spectral morphologies, the magnification is evaluated separately for every burst. 
The simulated energy distribution is therefore obtained from the collection of individual post-lensing energies rather than by multiplying the entire input energy distribution by a single gain factor.

Unlike the burst-rate calculation, no additional telescope detection threshold is applied when constructing the post-lensing energy function.
Here the purpose is to isolate the redistribution of burst energies caused by the lens within the rate-enhancement interval. 
The energy range used in the observational comparison lies well above the relevant detection threshold, so this selection does not affect the behavior examined here.

The resulting post-lensing energy functions are shown in Figure~\ref{fig:energy_function}. 
They provide the expected modification of the energy distribution if the observed high-rate episode were dominated by plasma-lens magnification and can therefore be compared directly with the observed Phase~I and Phase~II energy functions.


\bibliography{sample701}{}
\bibliographystyle{aasjournalv7}



\end{document}